# The Brightness of Starlink Mini Satellites During Orbit-Raising

Anthony Mallama*[1], Richard E. Cole, Jay Respler[1],
Scott Harrington, Ron Lee and Aaron Worley

2024 May 19

[1] IAU - Centre for the Protection of Dark and Quiet Skies from
Satellite Constellation Interference

* Correspondence: anthony.mallama@gmail.com

Abstract

Observations of Starlink V2 Mini satellites during orbit-raising suggest that SpaceX applies brightness mitigation when they reach a height of 357 km. The mean apparent magnitudes for objects below that height threshold is 2.68 while the mean for those above is 6.46. When magnitudes are adjusted to a uniform distance of 1000 km the means are 4.58 and 7.52, respectively. The difference of 2.94 between distance-adjusted magnitudes above and below threshold implies that mitigation is 93% effective in reducing the brightness of orbit-raising spacecraft.

Orbit-raising Mini spacecraft have a smaller impact on astronomical observations than higher altitude on-station spacecraft because they are relatively few in number. They also spend less time traversing the sky and spend longer in the Earth's shadow. These low-altitude objects will be more out-of-focus in large telescopes such as the LSST which reduces their impact, too. However, they attract considerable public attention and airline pilots have reported them as Unidentified Aerial Phenomena.

1. Introduction

Most Starlink spacecraft are injected into orbit at heights around 300 km. Then they ascend to on-station altitudes above 500 km under their own power. The brightness of second-generation (V2) Mini satellites during this orbit-raising phase is characterized in this paper.

Section 2 provides background information and summarizes a study of orbit-raising Starlink satellites of the first-generation. Section 3 describes how magnitudes were determined for this research. Section 4 characterizes the brightness of orbit-raising spacecraft. Section 5 discusses the impact of orbit-raising Mini satellites on astronomy. Section 6 presents our conclusions.

2. Background

Starlink satellites are a concern for astronomers because their brightness interferes with observation of the night sky (Barentine et al. 2023 and Mallama and Young 2021). In order to address this problem, SpaceX is using brightness mitigation-techniques to dim them. Mallama et al. (2023) evaluated the effectiveness of these efforts especially for on-station satellites.

SpaceX dimmed their first-generation Starlinks during orbit-raising by using a *knife-edge* configuration for the VisorSat and V1.5 (post-VisorSat) models. That orientation rolled the spacecraft to place the Sun in the plane of their flat surfaces. Mallama and Respler (2023) found that this satellite attitude made them 90% fainter than those without mitigation when magnitudes are adjusted to a uniform distance. This dimming is shown by the blue symbols and lines in Figure 1.

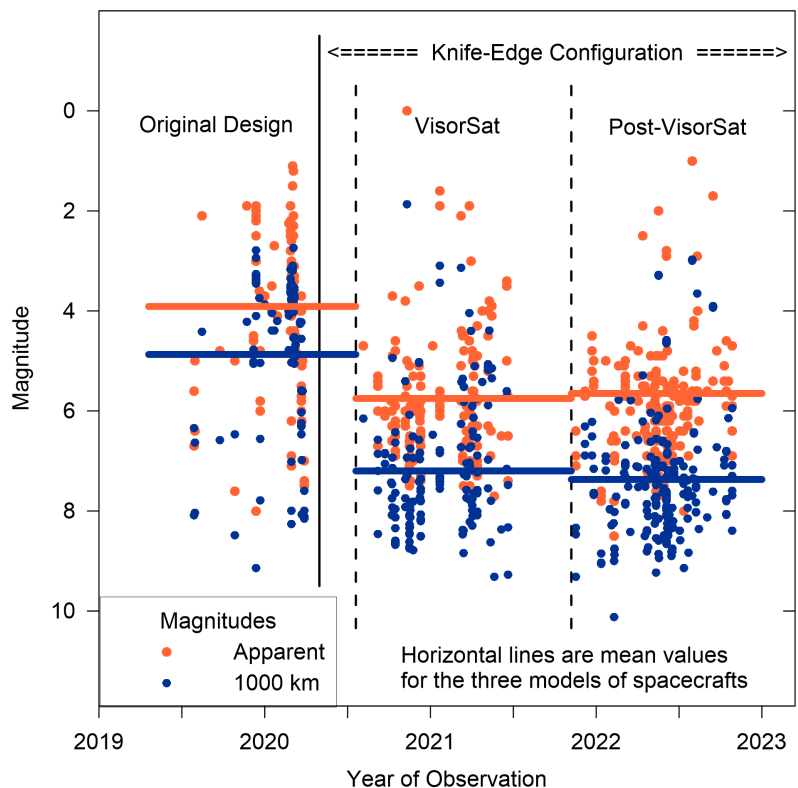

*Figure 1. Starlink Gen 1 satellites in orbit-raising were fainter after brightness mitigation began in 2020. Credit: Mallama and Respler (2023).*

SpaceX informed our team in July 2023 that they were mitigating the brightness of second-generation Mini satellites too. In this case, they were off-pointing the solar arrays to make them fainter from the ground.

3. Observations

The magnitudes analyzed in this study were recorded using electronic and visual methods. The former were obtained at the MMT9 robotic observatory (Karpov et al. 2015 and Beskin et al. (2017). The hardware of MMT9 consists of nine 71 mm diameter f/1.2 lenses and 2160 x 2560 sCMOS sensors. MMT9 magnitudes are within 0.1 of the V-band, based on information in a private communication from S. Karpov as discussed by Mallama (2021). We collected apparent magnitudes from the MMT9 on-line database along with ranges and phase angles.

The visual observation method, where spacecraft brightness is determined by comparison to nearby reference stars, results in magnitudes that approximate the V-band. The angular proximity between satellites and stellar objects accommodates variations in sky transparency and sky brightness. This method of observing is described in more detail by Mallama (2022).

Analysis required sorting the satellites into those that were orbit-raising and those already on-station. We used [graphs](#) of height versus time generated by J. McDowell to make that assessment. Satellites below 250 km were flagged in order to exclude de-orbiting spacecraft from the study. The upper height limit for orbit-raising satellites is taken to be 475 km because on-station orbit heights begin at about 480 km.

The distributions of apparent magnitudes for orbit-raising and on-station spacecraft are illustrated in Figure 2. That for orbit-raising satellites is strongly bimodal with peaks at magnitude 6 and 2. The fainter peak is similar to the distribution for on-station spacecraft.

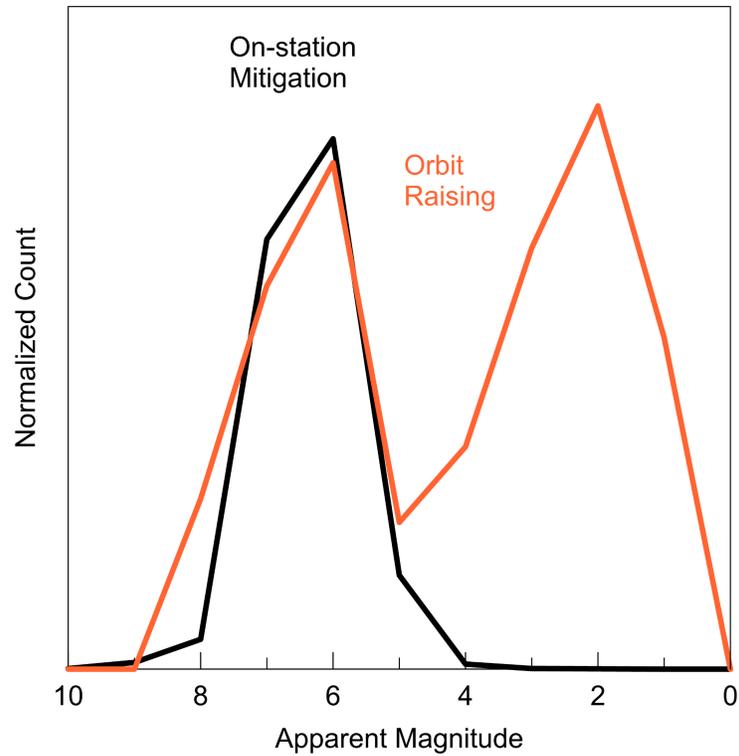

*Figure 2. Apparent magnitudes reveal a bimodal distribution for orbit-raising spacecraft.*

4. Brightness characterization

We investigated two relationships which might account for the bimodal brightness distribution shown in Figure 2. The first was brightness versus time since launch. Several observers have noted that Mini satellites are especially luminous during their first few days in orbit. For example, R. Lee suggests that a transition takes place about [2 days after launch](). Figure 3 reveals that Mini satellites do become fainter with time. The apparent magnitudes plotted there are brighter than those distance-adjusted to 1000 km because the ranges of low-altitude spacecraft are generally less than that uniform distance. Meanwhile, the linear fit to 1000-km magnitudes is less steep than that of apparent mags because they account for the increase of range with satellite height as time passes.

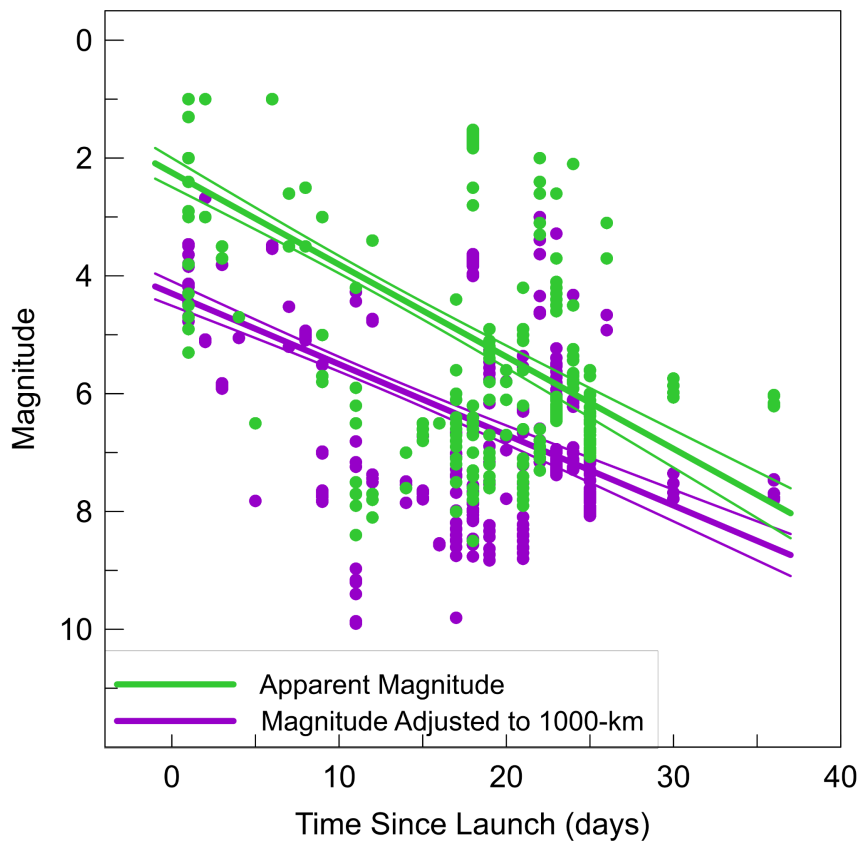

*Figure 3. Magnitudes of Mini satellites as a function of time since launch. Thin lines are confidence intervals.*

We also investigated satellite height as a possible cause for the bimodal brightness distribution of orbit-raising spacecraft. The 1000-km magnitudes plotted in Figure 4 reveal that satellites above a threshold height of 357 km are generally fainter than those below. The average of magnitudes above and below are 7.52 and 4.58, respectively. The difference of 2.94 magnitudes corresponds to 93% dimming which exceeds the value found by Mallama and Respler (2023) for first-generation satellites in knife-edge configuration.

Height, rather than time, is probably the root cause of the bimodal distribution. Satellites below the threshold height are subject to relatively high atmospheric drag. So, the spacecraft may be configured to reduce drag in addition to dimming them.

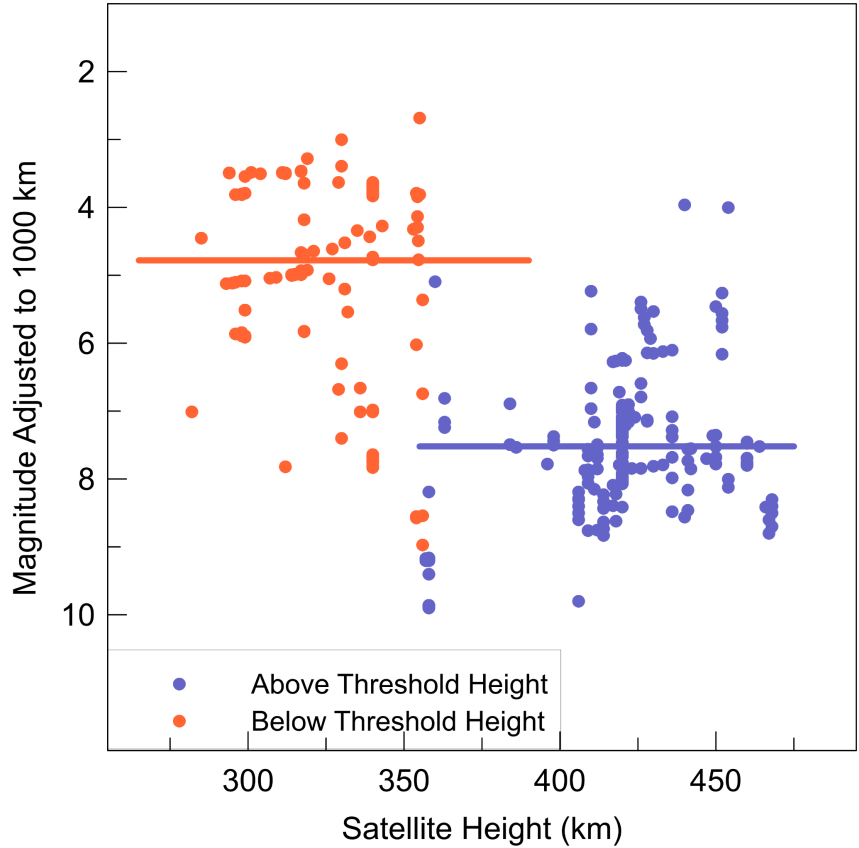

*Figure 4. Mini satellites above 357 km are generally fainter than those below. This plot contains 580 data points but many of them are overplotted, so it appears less. The standard deviations of the means are about 0.1 magnitude.*

5. Impact on astronomical research and beyond

Orbit-raising spacecraft are among the most luminous, however they only remain below the mitigation threshold height where they are bright for short durations. So, the number of unmitigated orbit-raising satellites at any given time is usually between 0 (if there are no very recent launches) and 23 (the greatest number of Minis orbited in one launch).

Besides their small numbers, there is another aspect that reduces the adverse impact of orbit-raising spacecraft on astronomy. Tyson et al. (2020) pointed out that satellites at lower heights are more out-of-focus in large aperture telescopes than those higher up. With out-of-focus objects, the total brightness is spread over more pixels, making the brightness per pixel less that what would happen with the same brightness at higher altitudes. Thus, the damage caused by the streak from a low-altitude satellite can be less serious than that due to a high-altitude spacecraft. Finally, orbit-raising satellites spend more time eclipsed by the Earth's shadow as compared to on-station spacecraft.

Despite their relatively low impact on astronomy, orbit-raising satellites below the threshold height attract a great deal of attention. Many sky watchers have reported seeing bright trains of newly launched Starlink satellites marching conspicuously across the sky. Unmitigated spacecraft traveling close together occasionally flare to magnitude 0 and brighter. Such eye-catching events are noticed by the general public and some have been covered by local news media.

Finally, airline pilots witnessing a bright train of low-altitude Starlink spacecraft reported them as Unidentified Aerial Phenomena. Buettner et al (2024) studied this extreme case where the satellites reached magnitude -4 or -5.

6. Conclusions

The brightness for Mini satellites during orbit-raising is characterized. The magnitude distribution of these spacecraft is bimodal. Our analysis suggests that SpaceX applies brightness mitigation to the satellites when they reach a threshold height of 357 km. The mean apparent magnitudes for objects below that threshold is 2.68, while the mean of magnitudes adjusted to a uniform distance of 1000 km is 4.58. The corresponding means for satellites above the threshold are 6.46 and 7.52. The difference of 2.94 magnitudes between distance-adjusted magnitudes above and below threshold implies that mitigation is 93% effective in dimming orbit-raising spacecraft.

Low-altitude Starlink Mini spacecraft have a smaller impact on astronomical observations than higher altitude on-station spacecraft because they are relatively few in number. These objects are also more out-of-focus in large aperture telescopes which further reduces their impact. They spend less time traversing the sky and spend longer in Earth's shadow, too. However, bright orbit-raising spacecraft attract public attention and have been reported as Unidentified Aerial Phenomena.

Acknowledgements

We thank the staff of the MMT9 robotic observatory for making their data available. The Heavens-Above.com web-site was used to plan observations. The Stellarium planetarium program and the Orbitron app were employed to process observations. We appreciate the internal review of this paper by the IAU - Centre for the Protection of Dark and Quiet Skies from Satellite Constellation Interference which helped to improve it.